\documentclass{revtex4}
\usepackage{epsfig}

\newcommand{\subpm}{{\!\raisebox{-2pt}{\tiny$+\!-$}}}
\begin{document}

\title{Decoherence due to telegraph and $1/f$ noise in Josephson qubits}

\author{E. Paladino,  A. Mastellone, A. D'Arrigo, and G. Falci}

\affiliation{Dipartimento di Metodologie Fisiche e Chimiche per l'Ingegneria 
(DMFCI) \\
MATIS, Istituto Nazionale per la Fisica della Materia (INFM) \\ 
Viale Andrea Doria 6, \\
95125 Catania, Italy\\
E-mail: epaladino@dmfci.unict.it}

\begin{abstract}
We study decoherence due to random telegraph and $1/f$ noise in Josephson qubits. We illustrate differences between 
gaussian and non gaussian effects at different 
working points and for different protocols.
Features of  the intrinsically 
non-gaussian and non-Markovian low-frequency noise 
may explain the rich physics observed in the spectroscopy
and the dynamics of charge based devices.
\end{abstract}

\maketitle

\section{Introduction}

Considerable progress 
has been recently achieved
in implementing two-state
systems using superconducting 
nanocircuits.
Time-resolved coherent oscillations have been measured in
Josephson qubits\cite{kn:Nakamura1,kn:vion,kn:Duty}
and signatures of the entanglement of coupled charge qubits have been
observed\cite{kn:two-qubit}.
Limitations in the performances arise from 
various noise sources\cite{kn:rmp,DPG,PRL,kn:galperin,kn:makhlin,kn:averin04},
often material and device dependent.
Recently evidence of 
noise due to individual impurities behaving as bistable fluctuators (BF) 
has been observed both in spectroscopy and in time resolved 
dynamical evolution\cite{Duty}. 
Sets of BFs are responsible for low frequency $1/f$ 
noise\cite{kn:weissman,kn:zorin}, which is 
the dominant dephasing mechanism common (to different extent) to all
solid state implementations.
A variety of behaviors, ranging from broadening 
due to a slow environment\cite{kn:Nakamura1,kn:vion}  
to relaxation limited 
decoherence\cite{kn:Duty}
 has been also observed, this phenomenology being strongly 
dependent on the particular device 
and on details of the protocol\cite{kn:vion,nakamura-echo,Varenna}. 
Explanation of this rich 
physics is beyond  phenomenological theories 
describing the environment as a suitable set of harmonic 
oscillators\cite{kn:rmp,kn:makhlin,kn:averin04}. Since the physical sources 
of noise are discrete in nature, attention has been devoted 
to environments made of collections of BFs\cite{PRL,DPG,kn:galperin}. 
In this work we will compare effects of gaussian and non-gaussian noise,
and show results indicating that discrete noise models potentially 
explain the experimental features, 
the main open 
question being a characterization beyond phenomenology 
of the physics of the noise sources\cite{kn:galperin}. 

\begin{figure}[t]
\centerline{
\epsfxsize=3in\epsfysize=1.43in\epsfbox{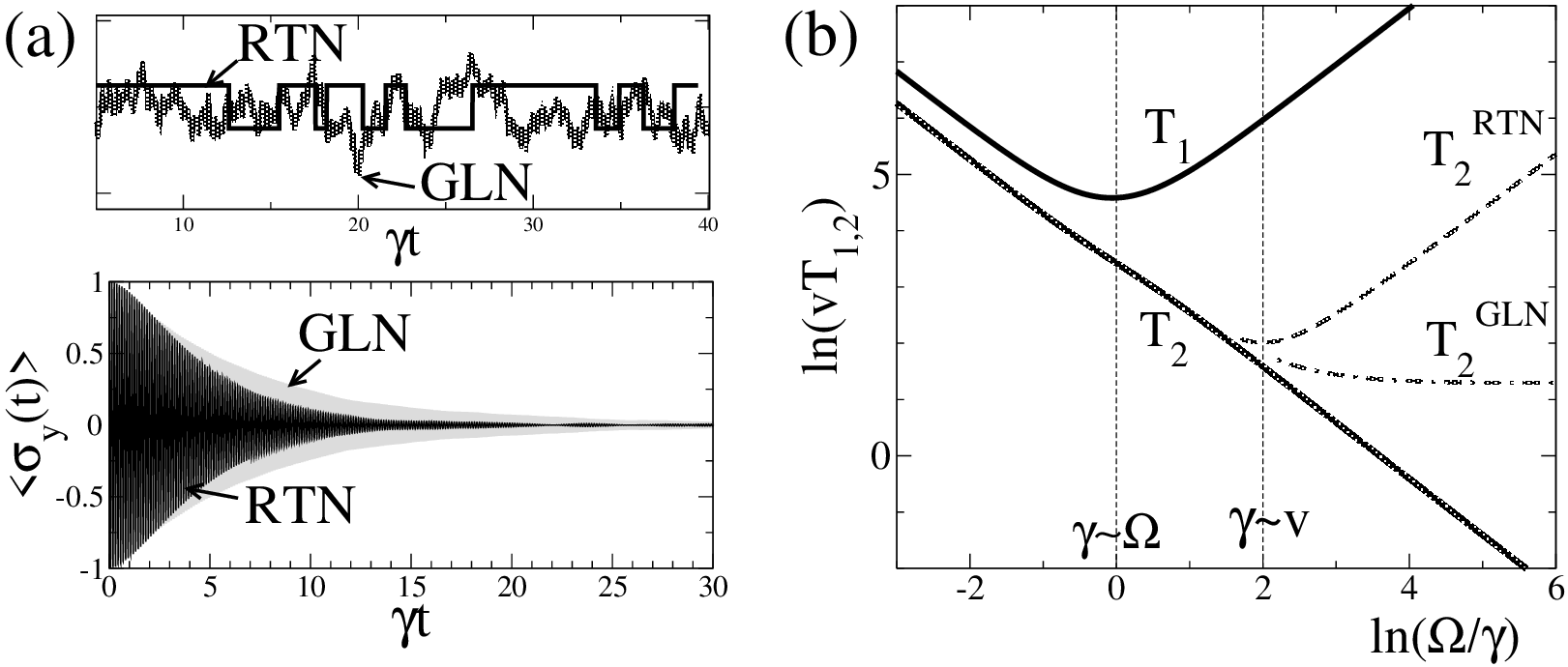}
\epsfxsize=1.6in\epsfysize=1.43in\epsfbox{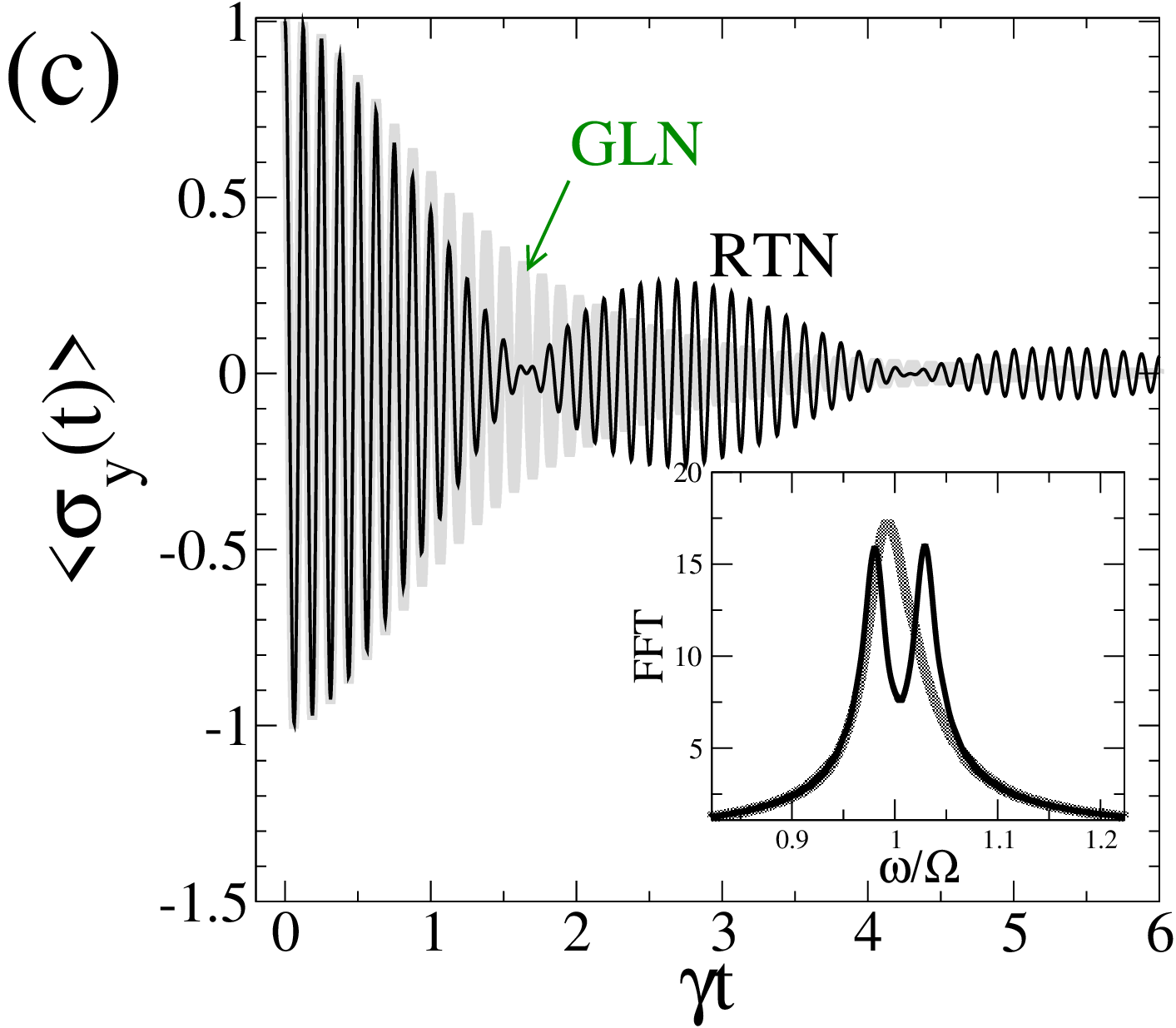}
}
\caption{(a) Upper panel: a 
sample of $\xi(t)$ for RTN and for GLN with
the same $S(\omega)$. Lower panel
$\langle \sigma_y \rangle$ for a BF with $v/\Omega=0.2$, 
$\gamma/\Omega=0.02$, has almost the same effect as GLN
at $\theta\approx\pi/2$.
(b) Relaxation and dephasing times by weak coupling theory (solid lines)
and behavior for slow RTN and GLN (dashed);
(c) Same set as in (a) for $\theta= 5\pi/12$ with 
RTN and GLN.  In the inset the 
spectral lines for RTN, showing two peaks and for GLN.
\label{fig:rates0}}
\end{figure}

Let us consider  the Hamiltonian 
$H =  H_Q - {v \over 2} \xi \sigma_z$, representing a 
qubit ($H_Q \, = - {1 \over 2} \vec{\Omega} \cdot \vec{\sigma}$) anisotropically\cite{kn:rmp} 
coupled to a noise source, described by $\xi$. Sensitivity to noise can be 
modulated by tuning the operating point, i.e. the angle 
$\theta$ between $\hat{z}$ and $\vec{\Omega}$, the qubit 
splitting $\Omega$ being also tunable. For classical noise 
$\xi \equiv \xi(t)$ is a stochastic process, whereas for 
quantum noise $\xi$ is an operator of the environment.
Decoherence results from the sensitivity of the qubit to the 
environment. For instance if coupling is 
weak, relaxation and dephasing rates are 
$T_1^{-1} = \sin^2\theta \, S(\Omega)/2$ and 
$T_2^{-1} = (2 T_1)^{-1} + T_2^{\prime \,\,-1}$, where 
$T_2^{\prime \,\, -1} = \cos^2\theta \, S(0)/2$ is the 
adiabatic rate, responsible for secular 
broadening\cite{kn:slichter}.
The power spectrum
$S(\omega)= v^2 \langle \xi \xi \rangle_\omega$ appears,
therefore the qubit is able to ``measure'' statistical 
properties of the environment at the level 
of two point correlations. 
In this regime 
the qubit is not sensitive to other details, for 
instance whether a certain Lorentzian line shape is due 
to a bistable fluctuator (BF) giving rise to 
Random Telegraph Noise (RTN) or to a continuous 
Gaussian process with {\em the same} Lorentzian 
Noise (GLN) spectrum (see Fig.~\ref{fig:rates0}.a). 
In solid state devices $T_2^{\prime \,\, -1}$  may be
very large thus invalidating  the weak 
coupling theory. This is the case of a sufficiently 
slow environment, no matter how small is $v / \Omega$.  
Consider for instance the Lorentzian spectrum
$S(\omega) = {v^2 \over 2} \, 
\frac{\gamma}{(\gamma^2 + \omega^2)}$. Rates in 
Fig.~\ref{fig:rates0}.b show that 
$T_2^{\prime \,\, -1} \sim v^2/\gamma$,
diverges for $\gamma \to 0$. The technical 
reason of this failure\cite{kn:cohen} 
is that the weak coupling expansion
parameter is $g \propto v/\gamma$. 
The high-frequency cutoff sets 
the characteristic time scale $\gamma^{-1}$ of
the noise, thus problems are encountered for slow noise, 
$g>1$, since the qubit becomes 
sensitive to details of the dynamics of the environment.
For instance it will distinguish RTN and GLN having
the same $S(\omega)$.
Indeed suppose we average the signal 
from an ensemble of experiments identical, apart for the 
uncontrolled preparation of the environment.
The effect of RTN due to a single BF is to determine 
two angular frequencies for the qubit, $\Omega$ and
$\Omega^\prime=\Omega [(v/\Omega+ \cos \theta)^2 + \sin^2 \theta]^{1/2}$,  
and the signal will show 
beats at the frequency $\Omega^\prime-\Omega$. 
If the BF is very slow, switches between the two frequencies 
produce decay with $T_{1,2} \sim \gamma^{-1}$. 
In this regime we can identify 
$g=(\Omega^\prime-\Omega)/\gamma$ from 
the condition\footnote{Notice that in the regime $v \ll \Omega$, $g$ is
approximated by 
$g \approx \cos \theta \, v/\gamma + \sin^2\theta \, v^2/(2 \gamma \Omega)$. 
Thus at $\theta=\pi/2$ it can be translated 
in $g \approx S(0)/\Omega$~\cite{kn:averin04} and the 
condition $g < 1$ simply means that broadening as given by 
the weak coupling theory should be smaller than the qubit 
splitting.}  for beats to be observable, $g>1$.
Instead for slow GLN\cite{kn:averin04}, a standard model
in NMR\cite{kn:slichter}, decay 
is determined by uncertainty in the 
preparation of the environment, an effect analogous to the 
``rigid lattice line breadth''\cite{kn:slichter}. 
By averaging the phase 
over a static 
distribution of the effective bias 
$\varepsilon_0 = v \xi_0$,
we find 
the decay of the qubit coherences 
$\Gamma(t) = - \ln |\rho_\subpm (t)|  = 
\ln \langle 
e^{i t 
(\Omega^2 + 2 \Omega \varepsilon_0 \cos \theta + \varepsilon_0^2)^{1/2} } \rangle$. The distribution of $\varepsilon_0$ 
is gaussian for GLN,
the standard deviation being 
$\sigma^2_{\!\varepsilon} = \int {d \omega \over 2 \pi} \, S(\omega) = 
\overline{v^2}/4$.
The decay depends on $\theta$
and it is slower for $\theta = \pi/2$ where for $v \ll \Omega$ a simple 
integration\footnote{The result gives the short 
time behavior found in Refs.\cite{kn:makhlin,kn:averin04} with a much more complete analysis.}
yields
$\Gamma(t) = -{1 \over 2} \ln\left(
1 + i \sigma^2_{\!\varepsilon} t/(2\Omega) \right)$.
For $\theta = 0$ the full dynamic problem can be solved
and $\Gamma(t) = (v/2 \gamma) (\gamma t - 1 + \mathrm{e}^{- \gamma t}) \approx (v t)^2/8$. In both cases the
signal amplitude is strongly reduced already at times 
$t \sim v^{-1} \ll \gamma^{-1}$ giving an apparent decay 
time $T_2^{*}$
independent on $\gamma$ and finite. 
Differences between slow RTN and GLN are also apparent in 
the line-shapes (see Fig.~\ref{fig:rates0}.c).

\section{Qubit dynamics in the presence of impurities}
\begin{figure}[t]
\centerline{\epsfxsize=1.9in\epsfysize=1.56in\epsfbox{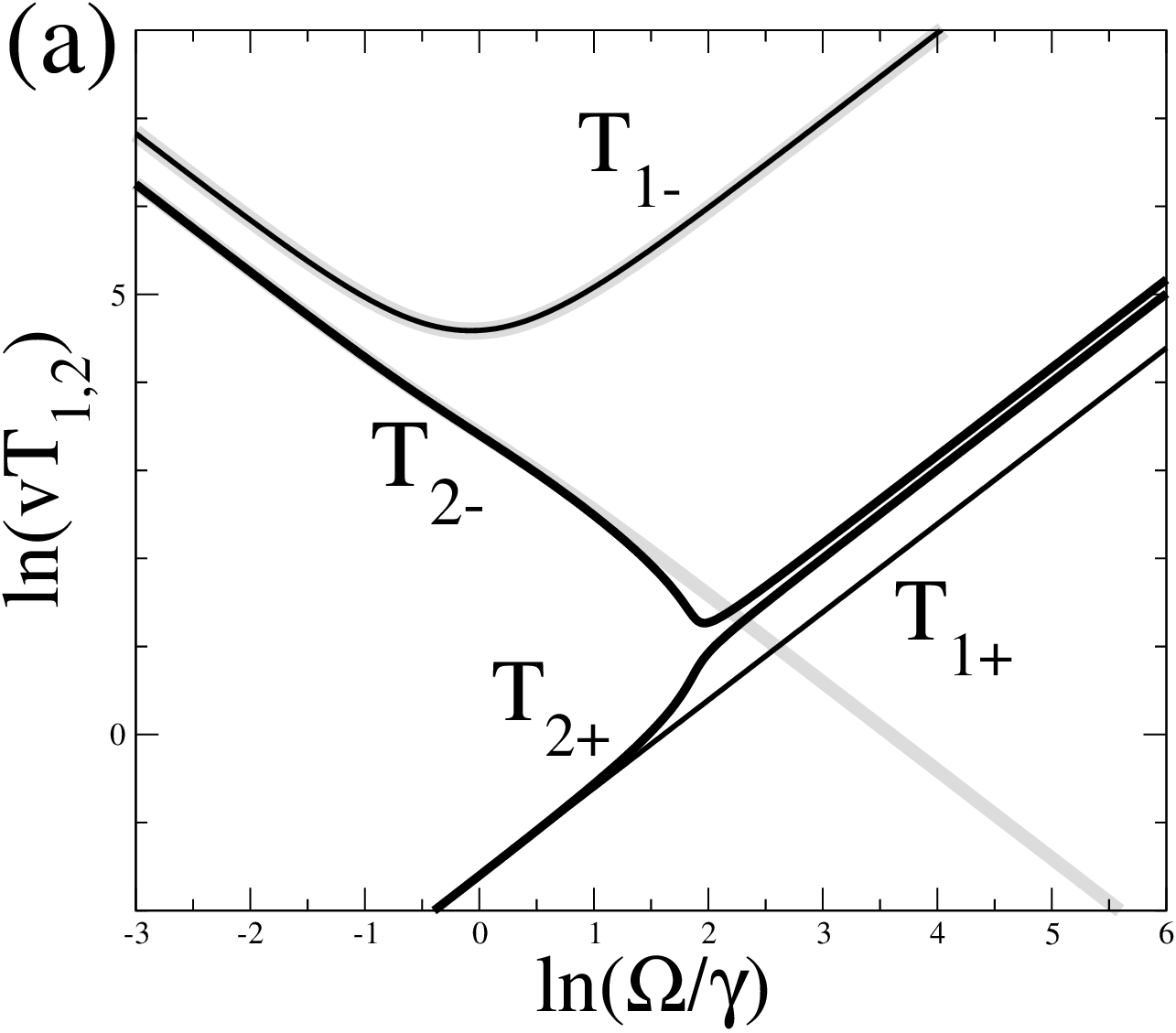}
\qquad
\epsfxsize=1.8in\epsfysize=1.56in\epsfbox{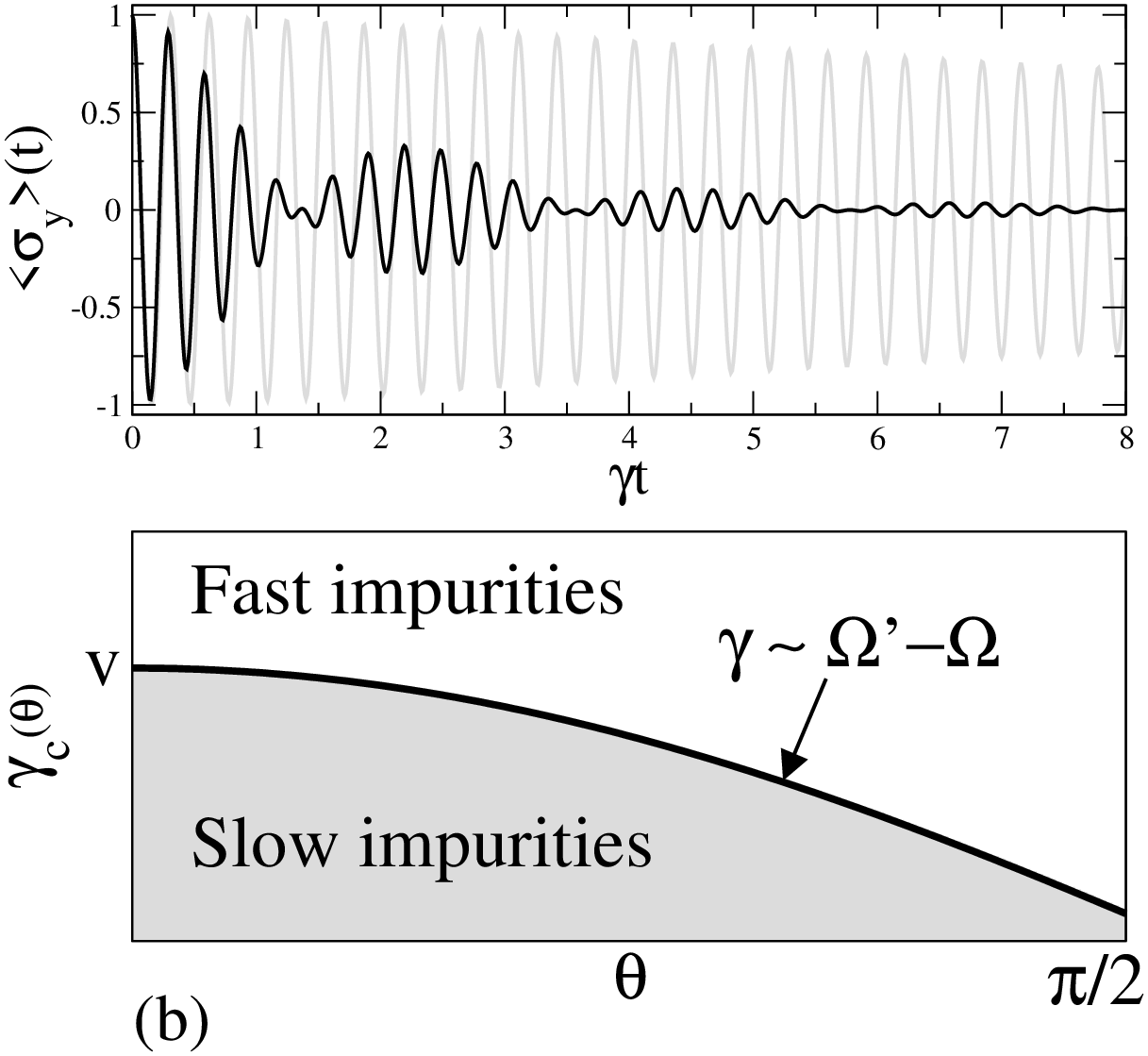}}
\caption{(a) Calculated relaxation (thin solid lines) 
and dephasing (thick solid) times versus 
$\Omega/\gamma$ at $\theta=\pi/4$ for 
decoherence due to an impurity with
$v/\Omega=0.2$, $ \overline{\delta p}=0.08$. 
Grey lines are the weak coupling result.
(b) Bottom: The crossover condition $\gamma_c(\theta)$. Top: 
$\langle \sigma_y \rangle$ for the same BF $v/\Omega=0.2$, 
$\gamma/\Omega=0.05$, tuned from the 
weak ($\theta=\pi/2$, gray) to the strong ($\theta=\pi/4$, 
black) coupling regime: the appearance 
of beats can be modulated with the external 
bias.
\label{fig:rates}}
\end{figure}

Decoherence due to discrete noise sources 
has been  studied in Ref.\cite{PRL,DPG} where the fluctuator
has been modeled by a quantum impurity described by the
Fano Anderson model. 
Rather than study higher orders\cite{kn:makhlin} 
of the perturbative expansion in $v$, 
a different strategy has been used, namely one may write a master equation 
(ME) for a four-level system system including
the impurity level, which is traced out at 
the end of the calculation. Thus fast components of 
the environment are effectively treated at the ME level,
whereas slow fluctuations are accounted for 
nonperturbatively. In practice one has at most to 
diagonalize a simple $8 \times 8$ Lindblad map which 
implies that the qubit dynamics is characterized by 
combining exponentials showing {\em two} relaxation times
$T_{1 \pm}$ 
and  {\em two}  dephasing times $T_{2 \pm}$
(Fig.~\ref{fig:rates}.a).
If $g \ll 1$ the dominant rates for 
$t \gg \gamma^{-1}$ coincide with the weak coupling ones (and the classical 
GLN ones) and describe 
homogeneous broadening.
Instead for impurities with $g \gg 1$, both $T_{2 \pm}^{-1}$
are important since dephasing reflects the 
bistable nature of the slow environment. Relaxation is not 
so sensitive: for $v \ll \Omega$, the dominant rate 
is the weak coupling one the other rate being 
$T_{1+}^{-1} \approx \gamma$. 
Rates have the form 
$T_{2 \pm}^{-1}+ i \delta \Omega/2 = \gamma(1 \pm \alpha)/2$ and 
$T_{1 \pm}^{-1}=\gamma(1 \pm \alpha_r)/2$, 
$\delta \Omega$ being the splitting between the two 
spectroscopic peaks induced by the bistable impurity\cite{DPG}.
They can be found analytically in the full adiabatic 
regime $\gamma < \Omega$, which includes the crossover 
region\cite{DPG}. In the stochastic limit for the BF
dynamics $\alpha= \big[(\overline{\delta p} -i g)^2
+\cos^4(\delta \theta)(1- \overline{\delta p}^2)\big]^{1/2}$ and
$\alpha_r= \big[ 1- \sin^2(\delta \theta) (1- \overline{\delta p}^2)\big]^{1/2}$, where $\delta \theta$ is the induced angular 
splitting on the Bloch sphere.   
Features of the nature of the BF appear if 
$\delta \Omega > \gamma$, i.e. $\Im \alpha > 1$, this 
criterion for ``strong coupling'' 
depending on {\em both} $v/\gamma$ and the operating point
(see Fig.~\ref{fig:rates}.b).  
Strongly coupled impurities are non-gaussian, 
and determine non-exponential decay. Moreover their 
effect depends on their preparation, so they are
non-stationary and non-Markovian, determining 
memory effects. As a consequence decoherence depends on details of 
the protocols\cite{Varenna}, a critical feature for 
$1/f$ noise.

\section{1/f noise at optimal working point}
Ideally quantum protocols assume we can measure 
individual members of an ensemble of identical (meaning that preparation is controlled)  time evolutions of the qubit. 
The environment limits control of the preparation. At most one 
can recalibrate, for each individual member, the collective variable 
$\xi$. With this feedback
scheme error would be due solely to decoherence {\em during} time evolution. 
In actual experiments lack of control on the environment
determines additional defocusing of the signal, 
analogous to inhomogeneous 
broadening in NMR\cite{kn:slichter}, which depends on the statistics of the 
environment at appropriate low-frequencies\cite{Varenna}.

We model $1/f$ noise with a set of impurities switching
at rates $\gamma_i$.  
For charge based devices they 
are coupled with the qubit via the total polarizing charge, 
$v \hat{\xi} \equiv \sum_i v_i n_i$, where $n_i=0,1$.  
The standard assumption\cite{kn:weissman} of a 
distribution of $\gamma_i$ with 
$P(\gamma) \propto 1/ \gamma$ for 
$\gamma \in [\gamma_m,\gamma_M]$\cite{kn:weissman}
leads to the  spectrum
$S(\omega)=  \sum_i \, \frac{1}{2} \, v_i^2 \,  
(1- \overline{\delta p}^2) \; \frac{\gamma_i}{(\gamma_i^2 + \omega^2)}$
which is $1/f$ at frequencies  
$2 \pi f \in [\gamma_m,\gamma_M]$.
In other words, 
impurities with both large and small $g$ 
are present. 
Decoherence for this model 
has been studied for $\theta=0$,
where exact solutions are available even for 
non-linear environments~\cite{PRL}. The feedback 
protocol results not to be sensitive to very slow impurities,  
having $\gamma_i < {\gamma}^* \sim \overline{v}/10$, which
sets an effective intrinsic low-frequency cutoff ${\gamma}^*$. 
This effect is due to the nongaussianity of the environment, and
to the fact that BFs with $g_i \sim 1$ determine strong
decay. Instead BFs with $g_i \ll 1$ behave as an environment
of quantum harmonic oscillators, being sensitive only to 
the amplitude of noise $\propto n_d \overline{v^2}$. In absence of 
recalibration,  
inhomogeneous broadening corresponds to a proper 
averaging of the initial conditions of the BFs and  
one may prove that ${\gamma}^*$
moves to ${\gamma}^*\sim \min\{\overline{v}/10,t_m^{-1}\}$, where 
$t_m$ is the overall measurement time of 
the experiment and a large number of repetitions is assumed.

\begin{figure}[t]
 \centerline{
\epsfxsize=1.9in\epsfysize=1.510in\epsfbox{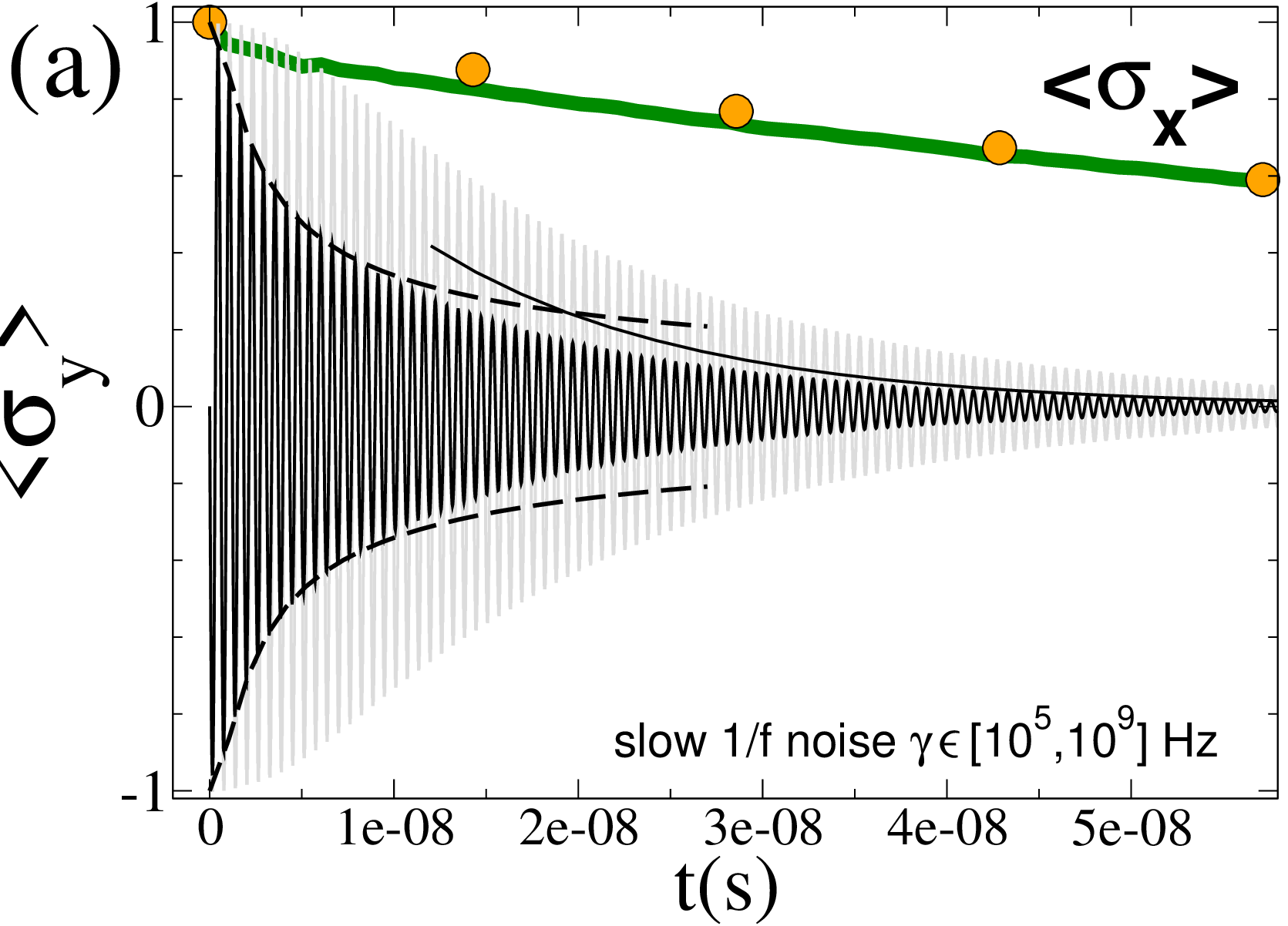}
\qquad
\epsfxsize=2.0in\epsfysize=1.51in\epsfbox{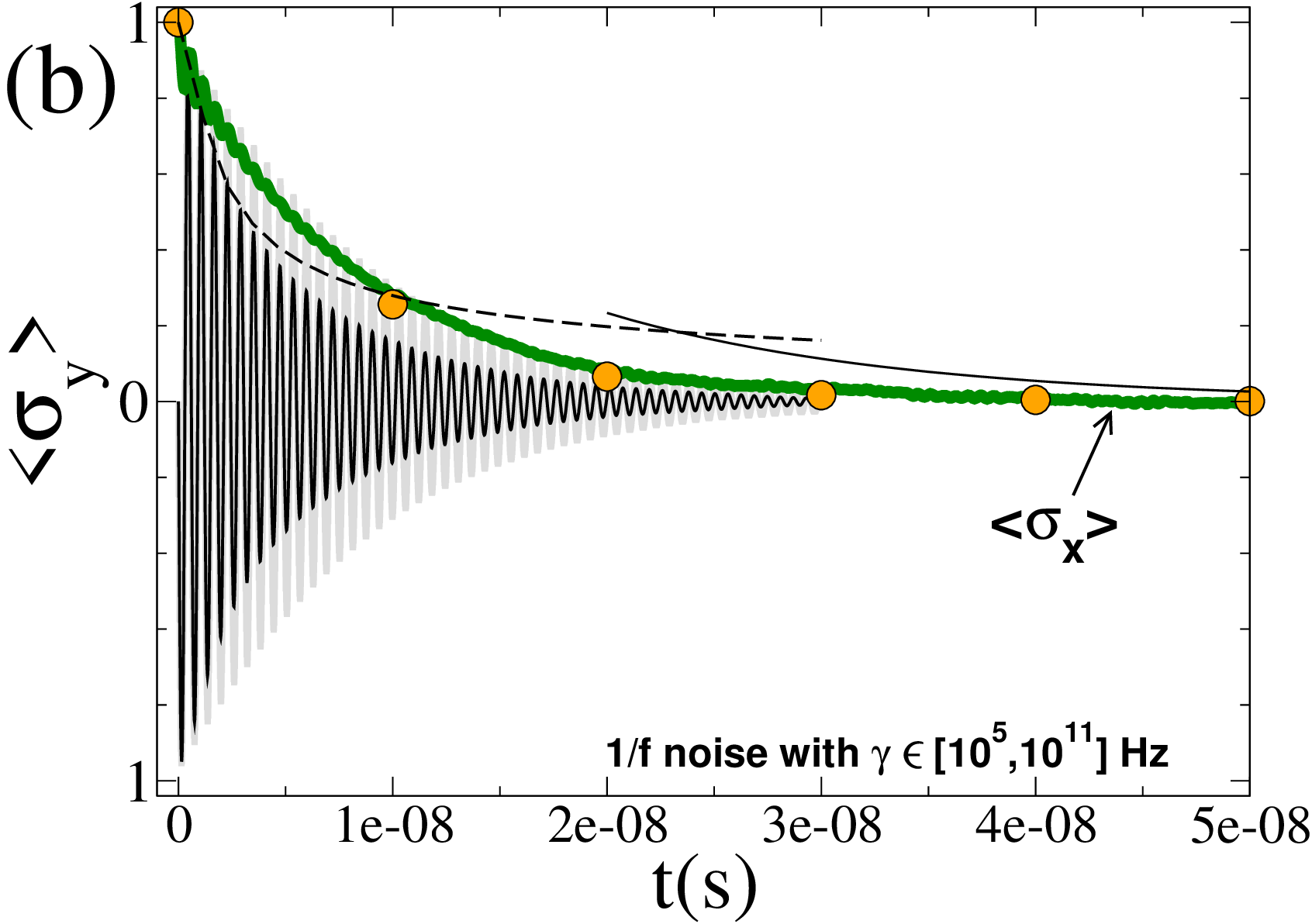}
}
\caption{Results of 
stochastic Schr{\"o}dinger equation, with up to 
$N_{BF}=1500$ BFs producing $1/f$ noise. Noise level
is appropriate to charge 
devices, 
$S(\omega)= 16 \pi  A E_C^2/\omega$ and $A \approx 10^{-6}$ and it is obtained 
with $n_d=250$ BFs per noise decade, with $\bar{v} = 0.02 \Omega$. 
Operating point is $\theta=\pi/2$ and $\Omega=10^{10}\,Hz$. In all the 
examples we show, relaxation coincides with the weak coupling result (thick solid line and 
dots).
(a) Adiabatic noise $\gamma_M= 10^9\,Hz$: the feedback procedure substantially decreases
the transverse  $\Gamma_\phi(t)$ (gray curves), whereas inhomogeneous 
broadening is well described by the rigid lattice breadth theory (dashed lines);
asymptotic behavior from Makhlin-Shnirman theory is also reported (thin line).
(b) Nonadiabatic noise $\gamma_M= 10^{11}\,Hz$: the main effect is decoherence 
during time evolution, determining $T_2 \approx T_1$.}
 \label{fig:2}
\end{figure}
The most effective implemented strategy\cite{kn:vion} 
for defeating $1/f$ is to tune optimally the working point, which
in our case means setting 
${\theta}=\pi/2$. Here the qubit splitting is less 
sensitive to bias fluctuations, or equivalently the lowest order 
adiabatic rate  $T_2^{\prime \,\, -1}$ vanishes. 
In other words, operating with $\theta$ modifies the 
parameters $g_i$ and the quantitative characterization 
of the spectrum in terms of ${\gamma}^*$, the main 
effect being that smaller $g_i$ means smaller effectiveness 
of noise. At first sight one expects that 
differences between a BF-$1/f$ environment and a set 
of oscillators with $1/f$ spectrum 
(G-$1/f$) decrease by operating at $\theta = \pi/2$, although
they do not necessarily disappear\cite{kn:galperin}.
Decoherence at the optimal point for a G-$1/f$ environment 
has been recently studied
by Makhlin and Shnirman\cite{kn:makhlin}, combining the 
adiabatic approximation and diagrammatic perturbation 
theory. 
Instead we study a BF-$1/f$ environment by 
solving numerically the stochastic Schr{\"o}dinger equation.
We checked that very slow BFs are ineffective, so we choose 
$\gamma_m = 10^5 \, Hz$.
We consider first an adiabatic environment, 
$\gamma_M  \ll \Omega$ (Fig.~\ref{fig:2}.a) studying 
relaxation via 
$\langle \sigma_x(t) \rangle$. The resulting rate reproduces 
the weak coupling result, in both protocols 
with and without recalibration of the environment. Instead
dephasing is much faster and recalibration is able to reduce
defocusing effects. Inhomogeneous broadening can be estimated
with the rigid lattice line breadth formula.~If one assumes $N_{FL}$ to be large enough that the initial
$\varepsilon_0 = \sum_{i=1}^{N_{FL}} v_i n_i(0)$ is gaussian 
distributed,  one finds
$\sigma^2_{\!\varepsilon} = 
\overline{v^2}N_{BF}/4= 
16 E_C A \ln{(\gamma_M/\gamma_m)}$, where physically 
$\gamma_m \equiv \gamma^* = 1/t_m$. This simple 
result accounts for the initial reduction of the signal 
amplitude, and coincides with the short-time behavior 
of the diagrammatic theory\cite{kn:makhlin}. Adding 
faster BFs to the environment increases relaxation more 
than dephasing. In our example (Fig.~\ref{fig:2}.b) 
decoherence is limited by relaxation $T_2 \approx T_1$, 
this latter being due to the fast part of the spectrum 
$\omega \sim \Omega$. 
Inhomogeneous broadening
does not reduce drastically the amplitude and, as expected,
dephasing is underestimated by adiabatic 
approaches~\cite{kn:makhlin}.

\begin{figure}[t]
\centerline{
\epsfxsize=2.3in \epsfysize=1.33in \epsfbox{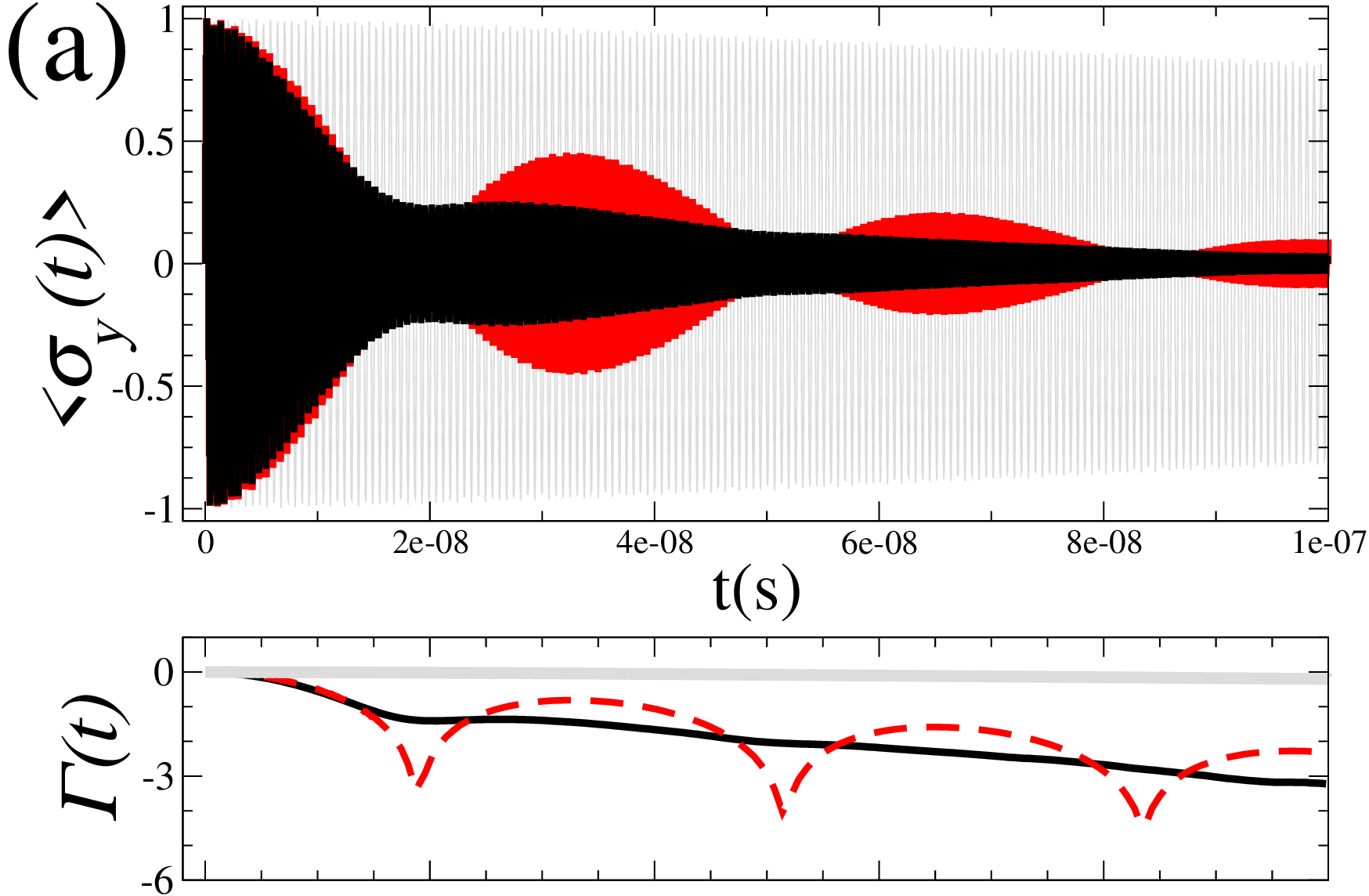}
\qquad
\epsfxsize=2.0in \epsfysize=1.33in \epsfbox{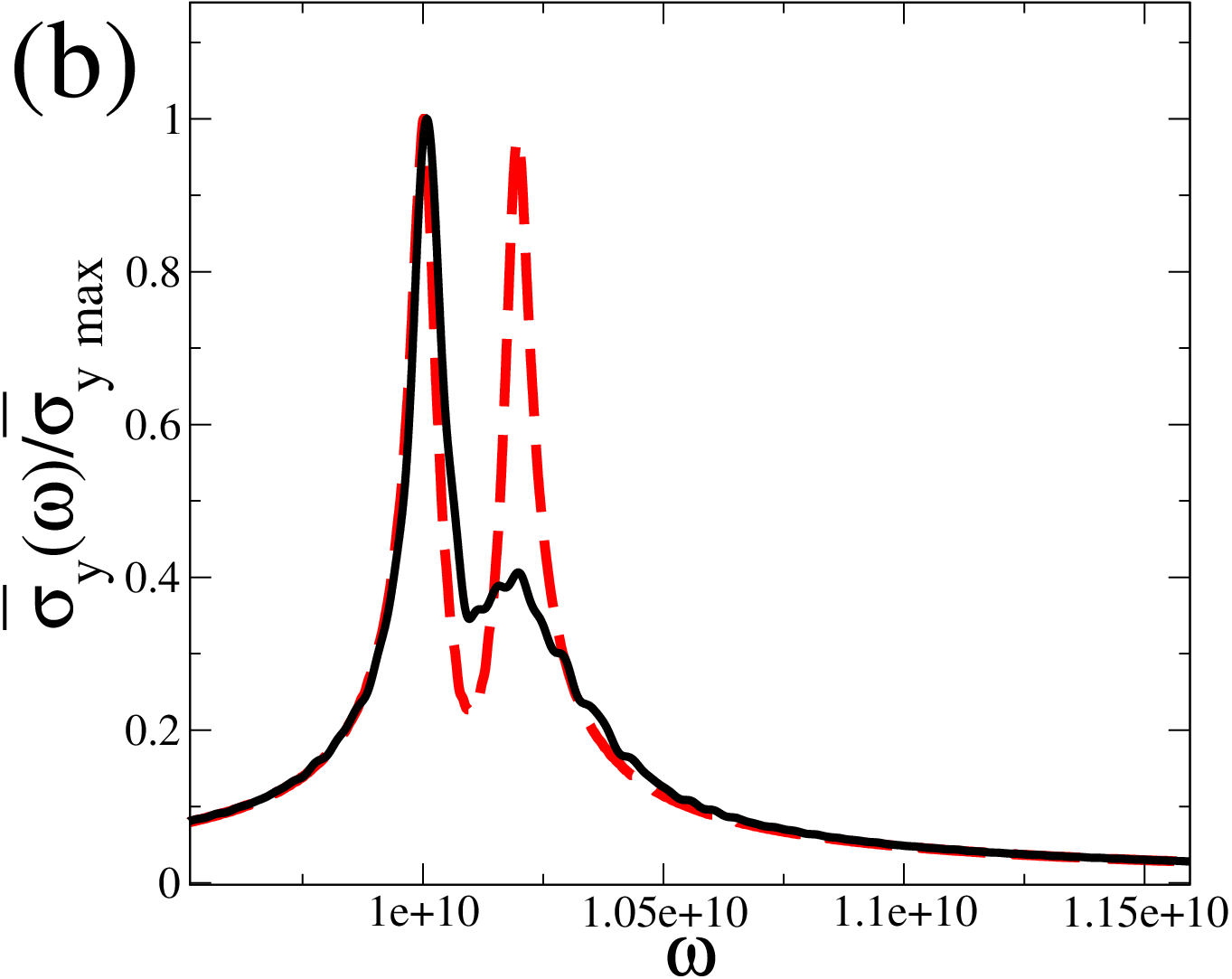}}
\caption{ (a)
$\langle \sigma_y \rangle$ at $\theta=\pi/2$, $\Omega=10^{10}$Hz. 
The effect of weak adiabatic $1/f$ noise (light gray line)
($\gamma \in [10^5, 10^9]$ Hz, uniform $v=0.002 \; \Omega$,  
$n_d=250$) is strongly enhanced by adding a single slow 
($\gamma/\Omega=0.01$)
more strongly coupled ($v_0/\Omega=0.2$) impurity (black line), which alone would give rise to beats 
(gray and dashed line). (b)
The latter two cases display a characteristic behavior of the Fourier transform, shown for 
the single impurity alone (dashed line) and for $1/f$ noise plus impurity (solid line).} 
 \label{fig:ft}
\end{figure}

Finally, we show that even a single impurity on a $1/f$ background may cause
a substantial reduction of the amplitude. 
This strongly poses the problem of reliability of charge based devices. 
Effects of 
realistic BF distributions were pointed out by Galperin et 
al.\cite{kn:galperin}. Using reasonable parameters, we obtain that indeed 
an additional BF has a substantial effect 
(see Fig.~\ref{fig:ft}), 
whose signature is an asymmetric Fourier transform double-peak,
which is similar to recent spectroscopy observations\cite{Duty}.

\end{document}